\documentclass[pra,tightenlines,12pt]{revtex4}
\usepackage{amssymb}
\usepackage{graphicx}
\usepackage{amsmath}
\usepackage{amsbsy}
\usepackage{amsthm}
\usepackage{bbm}
\usepackage{bm}
\newcommand{\id}{\mathbbm{1}}
\newcommand{\tr}{\textnormal{Tr}}
\newcommand{\ra}{\rangle}
\newcommand{\la}{\langle}

\begin{document}

\title{Two computable sets of multipartite entanglement measures}
\author{Beatrix C. Hiesmayr}
\affiliation{Faculty of Physics, University of Vienna,
Boltzmanngasse 5, 1090 Vienna, Austria}
\author{Marcus Huber}
\affiliation{Faculty of Physics, University of Vienna,
Boltzmanngasse 5, 1090 Vienna, Austria}
\author{Philipp Krammer}
\affiliation{Faculty of Physics, University of Vienna,
Boltzmanngasse 5, 1090 Vienna, Austria}

\begin{abstract}
We present two sets of computable entanglement measures for
multipartite systems where each subsystem can have different degrees
of freedom (so-called qudits). One set, called ``separability''
measure, reveals which of the subsystems are separable/entangled.
For that we have to extend the concept of $k$--separability for
multipartite systems to a novel unambiguous separability concept
which we call $\gamma_k$--separability. The second set of
entanglement measures reveals the ``kind'' of entanglement, i.e. if
it is bipartite, tripartite, \dots , $n$-partite entangled and is
denoted as the ``physical'' measure. We show how lower bounds on
both sets of measures can be obtained by the observation that any
entropy may be rewritten via operational expressions known as
$m$--concurrences. Moreover, for different classes of bipartite or
multipartite qudit systems we compute the bounds explicitly and
discover that they are often tight or equivalent to positive partial
transposition (PPT).\\
\\
Keywords: entanglement measure, multipartite qudit system, separability\\
PACS: 03.67.Mn
\end{abstract}

\maketitle


\section{Introduction}

Quantum entanglement is a fascinating property of quantum states
that has many important consequences for modern physics. It exhibits
aspects that are counter-intuitive to classical physics, like the
incompatibility with local realistic theories \cite{bell64,
clauser69}. For example it turned out that a symmetry violation in
particle physics, the $CP$ violation in mixing ($C$\dots charge
conjugation, $P$\dots parity), is incompatible with any local
realistic theory \cite{BGH4,H2}.

Furthermore entanglement is a highly useful resource for quantum information
tasks. Thus it makes quantum information theory a conceptually different theory than classical
information theory (for an overview see, e.g.,
Refs.~\cite{nielsen00, bouwmeester00, bertlmann02a}). The
characterization of entanglement is, however, a nontrivial
mathematical task and not at all completed (for an overview see,
e.g., Refs.~\cite{bruss02, horodecki07}). The first concepts were
derived for bipartite systems, which are the simplest systems that
can contain entanglement. Here many important results were obtained,
like the detection and quantification of entanglement for (pure and
mixed) bipartite qubits, which can be conclusively performed for any
states of such systems. In a finite dimensional Hilbert space the
most general quantum states one can think of are multipartite
arbitrary dimensional states, i.e. states that describe systems of
$n$ subsystems, where each subsystem is ascribed a finite
dimensional Hilbert space. Already the classification of
entanglement according to possible reversible quantum operations is
a nontrivial task, see e.g. Refs.~\cite{duer00a, verstraete02,
verstraete03} in this context.

There are different approaches to the quantification of multipartite
entanglement. A common method is to describe the same state with
different entanglement measures, e.g. in terms of bipartite cuts
\cite{coffman00} or with different entanglement measures according
to invariance classes under statistical local operations and
classical communication (SLOCC) \cite{verstraete03}. Another way is
to determine a global entanglement measure for the whole state
\cite{bennett00, kruszynska08, meyer02, teresi08, rigolin06,
somma04, barnum04, demkowicz-dobrzanski06}. Our approach picks up a
conception of entanglement that on the one hand differentiates
between possible entanglement between any parties sharing the state
and on the other hand sums up to a total global entanglement. In
this way we can both quantify the entanglement that any parties
share with each other, and the ``whole'' entanglement present in the
state.

This would also provide advantages for the description of quantum
communication protocols with multipartite entangled states (see,
e.g., Refs.~\cite{kempe99, scarani01, cerf02, kruszynska06}). Of
course this simple concept already calls for more than one
entanglement measure - for different tasks different entanglement
measures seem to be appropriate. We want to present two of such
possibilities that in our opinion seem to be good candidates, i.e.
one revealing the separability property, ``the separability
measure'', and the other one revealing different kinds of
entanglement, ``the physical measure''.
Further approaches to quantify multipartite entanglement can be found, e.g., in Refs.~\cite{bennett00, shi02, demkowicz-dobrzanski06} and higher dimensional generalizations of bipartite entanglement measures in \cite{uhlmann00, rungta01,
audenaert01, mintert05}.

The paper is organized as follows: In the first section we define
separability of multipartite systems and list the requirements for bipartite
entanglement measures. In Sec.~\ref{chentropy} we discuss entropies and
introduce the $m$--concurrence which enables computation of bounds
on entanglement of mixed states. The next Sec.~\ref{chmeasure}
introduces two measures, one for the partial separability and one
for what kind of entanglement is present. Then follows a section
with further instructive examples to which we applied the two
measures. In the appendix we give all proofs of the requirements for
these two measures.

\section{Basic definitions}

\subsection{A definition of \textit{partial} separability and the $\gamma_k$--separability}

In multipartite systems the notion of separability can be extended
in order to answer the question which particles are joint
inseparably. Throughout the paper we assume that partial traces of
the multipartite quantum system are only taken over physical
subsystems, i.e. over one or more particles. It means that possible
information which may result by tracing over certain degrees of
freedom of a certain particle/qudit is not taken into account.


A pure multipartite state $|\psi\rangle$  is called
$k$--\emph{separable} if it can be written as \cite{horodecki07}
\begin{eqnarray}
|\psi\rangle=|\phi_1\rangle\otimes|\phi_2\rangle\otimes\cdots\otimes|\phi_k\rangle
\,, \quad k \leq n \,,
\end{eqnarray}
where $n$ is the total number of particles. It is called fully
separable iff $k=n$, this is the natural generalization of the
separability of bipartite systems.  We
call a state $1$--separable or fully entangled iff $k=1$. This notation of full separability or entanglement can be generalized to mixed states in a straight forward way. If a pure state is not fully separable or fully entangled, it is called partially separable or $k$--separable.

The definition of partial separability for mixed states is more involved. One obvious possibility is the following: A mixed
state is called $k$--separable if there exists a decomposition that
satisfies \cite{horodecki07}
\begin{eqnarray} \label{horsep}
\sigma_{k-sep}=\sum_i\; p_i\;
\rho_i^1\otimes\rho_i^2\otimes\cdots\otimes\rho_i^k \,, \qquad
\textrm{with}\quad p_i\geq 0,\; \sum_i p_i=1\;,
\end{eqnarray} where the $\rho_i^j$s are states of some number of subsystems and can always be chosen to be pure. The terms in Eq.~\eqref{horsep} all have the same $k$, but it is in general not fixed which subsystems are contained in the states $\rho_i^j$.

For the argumentation in this paper we are interested to fix the subsystems involved in the states $\rho_i^j$
and therefore extend the $k$--separability definition to the so-called $\gamma_k$--separability. For this we introduce the
following notation:
\begin{eqnarray}
\gamma_k:=\{\{\beta_1\}|\{\beta_2\}|\cdots|\{\beta_k\}\} \,.
\end{eqnarray}
Here the sets $\{\beta_j\}$ represent subsystems, i.e. particles,
which are inseparably joined.
\begin{center}
\begin{tabular}{l}
\parbox[b]{12cm}{\begin{small}\textbf{Instructive example:} $|\psi\rangle=|0\rangle_1\otimes|0\rangle_2\otimes|\phi^+\rangle_{34}$ with
$|\phi^+\rangle=\frac{1}{\sqrt{2}}\{|0\rangle\otimes|0\rangle+|1\rangle\otimes|1\rangle\}$.
Here the number of particles is $n=4$ and the separability is a
$3$--separability with the substructure $\gamma_3=\{1|2|34\}$.\\
This state is obviously equivalent to
$\frac{1}{\sqrt{2}}\{|0000\rangle+|1010\rangle\}$ with the
substructure $\gamma_3=\{2|4|13\}$, here just the role of the first
and second subsystems are interchanged. Therefore, it is convenient
to reorder the subsystems of the state if necessary.
\end{small}}
\end{tabular}
\end{center}

Note that there is a difference between fully entangled and being
maximally entangled, which we define as a pure state where all
subsystems representing particles are in the maximally mixed state.
For example the state $|\phi^+\rangle\otimes|\phi^+\rangle$ is not
fully entangled, but according to the above definition maximally
entangled.

The extension of the $\gamma_k$ separability to mixed states is not
straightforward as an ambiguity can happen as we explain later in an
example.\\
\\
\textbf{Definition} of
$\gamma_k$--separability:\\
To every $\rho$ we associate a separability property, the set
$\gamma_k$, which is made up of $\{\beta_j\}$, i.e. sets of numbers
representing subsystems. A state $\rho$ is called
$\gamma_k$--separable iff there exists an unambiguous decomposition
with maximal $k$ into:
\begin{eqnarray}
\sigma_{\gamma_k-sep}=\sum_i\; p_i\;
\rho_i^{\{\beta_1\}}\otimes\rho_i^{\{\beta_2\}}\otimes\cdots\otimes\rho_i^{\{\beta_k\}}
\,, \qquad \textrm{with}\quad p_i\geq 0,\; \sum_i p_i=1\;.
\end{eqnarray}

The following instructive example shows the difference of the
$k$--separability and the $\gamma_k$--separability.
\begin{center}
\begin{tabular}{l}
\parbox[b]{12cm}{\begin{small}
Consider the generalized Smolin state
\cite{hiesmayr08a,HorodeckigeneralizedSmolin,Smolin}
\begin{eqnarray}\rho_{Smolin}=\frac{1}{2^n}\left(\mathbbm{1}+\sum c_i\;
\sigma_i^{\otimes n}\right)\;,\end{eqnarray} where $n$ is an even
number, $\sigma_i$ are the Pauli matrices and $c_i$ are real numbers
(see also the instructive example in
Sect.~\ref{separabilitymeasure}). This state can be decomposed into
bipartite pure states, i.e. the Bell states. For $n=4$ this would
correspond to $\gamma_k=\{12,34\}$. This however is not the proper
$\gamma_k$ separability as any other bipartite cut is also valid,
i.e. $\gamma_k=\{13,24\}$, $\gamma_k=\{14,23\}$. So the only
unambigous set of subsystems is: $\gamma_k=\{1234\}$. So in the
notion of $\gamma_k$--separability, the generalized Smolin states
are always completely inseparable ($\gamma_1$-separable), whereas in
the notion of $k$-separability they are
$\frac{n}{2}$-separable.\end{small}}
\end{tabular}
\end{center}

Both views are in a way justified: The $\gamma_k$-separability
reflects the fact that any further reduction (partial trace) of the
state yields a fully separable state (which is independent of
notion) and therefore the useful entanglement properties can only be
extracted if one uses all contained subsystems. Whereas the
$k$-separability reflects how many parties need to join together in
order to prepare the state using LOCC. Note again that for pure
states the $k$ in $k$-separability and the $k$ in $\gamma_k$ are
identical.

\begin{figure}
\begin{center}
\includegraphics[width=10cm, keepaspectratio=true]{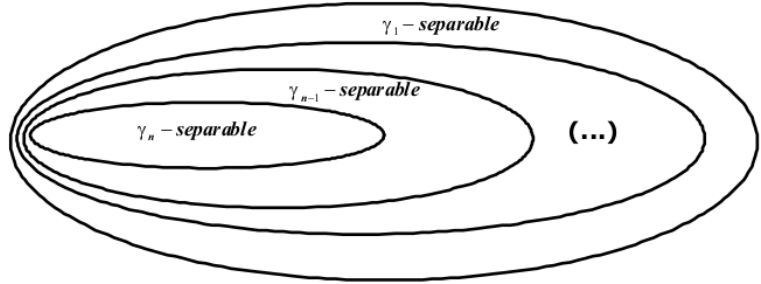}
\caption{Here the convexity of $\gamma_k$-separability is
visualized, i.e. any convex mixture of two $\gamma_k$--separable
states, e.g. $\gamma_{k_1}$ and $\gamma_{k_2}$, is either
$\gamma_{k_1}$-- or $\gamma_{k_2}$-- or $\gamma_k$--separable with
$k_1, k_2<k$.} \label{gammak}
\end{center}
\end{figure}

To sum up, $\gamma_k$-separability for pure states is an extension
of $k$-separability, it captures which subsystems are involved, and
for mixed states it captures an essential novel feature (cf. the
above example of the Smolin states) that would be missed by only
considering $k$-separability. Another important feature of the
$\gamma_k$--separability is the convexity in the sense that the
mixture of two $\gamma_k$--separable states, e.g. $\gamma_{k_1}$ and
$\gamma_{k_2}$, is either $\gamma_{k_1}$-- or $\gamma_{k_2}$-- or
$\gamma_k$--separable with $k_1, k_2<k$. This is visualized in
Fig.~\ref{gammak}.

One aim of this paper is to quantify entanglement and classify the $\gamma_k$--separability of a given state
which is done in Sect.~\ref{separabilitymeasure}.

\subsection{Proper properties for being entangled}

Now we investigate the question what properties a proper
entanglement measure should have. Let us first summarize the
conditions which are required for bipartite entanglement measures
$E(\rho)$ ($Sep$ is the set of all separable states)
\cite{bennett96b, bennett96, vedral97, vedral98}:
\begin{enumerate}
    \item[B1:] $E(\rho)>0\quad\forall\quad \rho\not\in Sep$
    \item[B2:] $E(\rho)=0\quad\forall\quad \rho\in Sep$
    \item[B3:] $E(\rho^{\otimes n})=n\;E(\rho)$ (Additivity)
    \item[B4:] $E(\lambda\, \rho_1 + (1-\lambda)\,\rho_2)\;\leq\;\lambda\, E(\rho_1) +
    (1-\lambda)\,E(\rho_2)$ (Convexity)
    \item[B5:] $E \left( U^A \otimes U^B \,\rho \, \left( U^A \otimes U^B \right) ^\dagger \right) = E(\rho)$ (Invariance under local unitary operations)
    \item[B6:] $\sum_i \tr \left( V_i \rho V_i^\dag \right) E \left( \frac{V_i \rho V_i^\dag}{\tr \left( V_i \rho V_i^\dag \right)} \right) \leq E(\rho)$ (Non-increasing on average under LOCC),\\
        where $V_i$ is a separable operator, i.e. of the local form $V_i := A_i \otimes B_i$\,.
\end{enumerate}

For multipartite systems we claim that there cannot be only a single entanglement
measure, since it could not correctly quantify the substructure of the
$k$--separability or the $\gamma_k$--separability and
simultaneously reveal which parts of the system are entangled in
which way with other parts.

Therefore we propose for multipartite systems a set of entanglement
measures $E_{\{\alpha_j\}}$ where the set $\{\alpha_j\}$ denotes
subsystems of the whole. As any bipartite system can be seen as a
substructure of a bigger system, clearly the same requirements as
for bipartite systems applies also to bipartite $E_{\{\alpha_j\}}$.
The same should hold true for any tripartite, four-partite, \ldots and so on, entanglement.
The sum over the whole set should constitute the total entanglement
\begin{eqnarray}
E_{tot}(\rho)=\sum_{j=2}^n\sum_{\{\alpha_j\}}E_{\{\alpha_j\}}(\rho)\;.
\end{eqnarray}
It is well known that the entanglement of a pure state can easily be
quantified by the entropy of its subsystems \cite{bennett96}. Possible
entropy measures are, e.g., the quantum version of Renyi's $\alpha$--entropies \cite{renyi60}:
\begin{eqnarray}\label{reny}
S_\alpha^q:=\frac{1}{1-\alpha}\log_q\text{Tr}(\rho^\alpha)
\end{eqnarray}
which for $\alpha\rightarrow 1$ equals the famous von Neumann
entropy. The logarithmic entropies have the advantage that they
imply additivity, for the general cases of probability
distributions, for which they were originally intended, as well as
for entanglement measures constructed out if it.

Another possibility are the linear entropies
\begin{eqnarray}\label{entropy}
S_{r}(\rho):=\frac{d^{r-1}}{d^{r-1}-1}(1-\text{Tr}(\rho^r))\;,
\end{eqnarray}
where $d$ is the dimension of $\rho$.

For any multipartite pure state $\rho=|\psi\rangle\langle\psi|$ one
can quantify the total entanglement by
\begin{eqnarray}
E_{tot}(\rho):=\sum_{s=1}^n S(\rho_s)\;,
\end{eqnarray}
where $\rho_s:=\text{Tr}_{\neg s}\rho$ denotes the reduced density
matrix of the respective subsystem $s$ and $S$ is any entropy
function. A standard method to generalized this measure for mixed
states $\rho=\sum_i p_i |\psi_i\rangle\langle\psi_i|$  is by
constructing the convex roof \cite{Uhlmann}
\begin{eqnarray}
E_{tot}(\rho):=\inf_{\{p_i,\psi_i\}}\sum_i p_i\sum_{s=1}^n
S(\rho_s^i)\;.
\end{eqnarray}
The $\psi_i$ are throughout the paper considered as normalized. In
general it is not known how to find the infimum, we will show in the
next section how with a simple algebraic trick operators can be
constructed which allow to compute bounds on the entanglement which
turn out to be tight in many cases.

\section{Entropy and $m$--Concurrence}\label{chentropy}
The $m$--concurrence was introduced in Ref.~\cite{HH2}. It can be efficiently used to compute bounds for the convex roof extension of the entanglement measures for mixed states. For pure multipartite
states it is a simple algebraic rewriting of the entropies of the
subsystems in terms of such operators. For the generally mixed
subsystems states one can via these operators obtain bounds on the
entanglement.

The definition we present here will be slightly modified to the
above cited works in oder to yield a simpler and more compact
notation. The linear entropy $S_{r=2}$, Eq.~\eqref{entropy}, of any
subsystem $s$ can be rewritten as a sum of terms named concurrences
in analogy to Hill and Wootters concurrence \cite{hill97,
Wootters98} and concurrences defined for bipartite systems of
arbitrary dimension \cite{rungta01, audenaert01}:
\begin{eqnarray}
S_2(\rho_s)&=&\frac{d}{d-1}(1-\text{Tr}(\rho_s^2))\nonumber\\
&=&\sum_{\alpha} C^2_{\textbf{s}\alpha}+\sum_{\alpha}\sum_{\beta}
C^2_{\textbf{s}\alpha\beta}+(\ldots)+\sum_{\alpha}\sum_{\beta}\cdots\sum_\omega
C^2_{\textbf{s}\alpha\beta\cdots\omega}\;.
\end{eqnarray}
This $m$--concurrences $C_{\{\alpha_m\}}^2$ containing $m$--indices
are obtained using $m$--flip operators in the following way:
\begin{eqnarray}\label{mconcurrence}
C_{\textbf{s}\alpha\beta\cdots\omega}^2:=\sum_{O_C}\left|\langle\psi|\underbrace{(A|\{i_n\}\rangle\langle\{i_n\}|\mathbbm{1}-B|\{i_n\}\rangle\langle\{i_n\}|AB)}_{O_C}|\psi^*\rangle\right|^2
\end{eqnarray}
where
\begin{eqnarray}
A&:=&\left(\sigma_{k_Kl_K}^{K\in\{\textbf{s}\alpha\beta\cdots\omega\}},\id^{K\notin\{\textbf{s}\alpha\beta\cdots\omega\}}\right)\nonumber\\
B&:=&\left(\sigma_{k_Kl_K}^{K=\textbf{s}},\id^{K\neq\textbf{s}}\right)
\end{eqnarray}
and
\begin{eqnarray}
\sum_{O_C}:=\sum_{k_K=0}^{d_K-1}\sum_{l_K>k_K}\sum_{\{i_n\}}\;.
\end{eqnarray}
Here $K$ denotes the respective subsystem and the flip operators are defined for a qudit system of dimension $d$ in the following way:
\begin{eqnarray}
&&\sigma_{kl}^{d\times
d}\;|k\rangle\;=\;|l\rangle,\quad\sigma_{kl}^{d\times
d}\;|l\rangle\;=\;|k\rangle\quad\textrm{and}\quad
\sigma_{kl}^{d\times d}\;|t\rangle\;=\;0\quad \forall\; t\neq k,l\;.
\end{eqnarray}
Note that these are the symmetric generalized Gell-Mann operators
(see, e.g., Refs.~\cite{kimura03, bertlmann08a}; generalized
Gell-Mann operators are the SU(N) generators). In order to obtain
Renyi's entropy we use the relationship between this entropy
\eqref{reny} and the linear entropy \eqref{entropy}
\begin{eqnarray}
S_\alpha^q(\rho)=\frac{1}{1-\alpha}\textbf{log}_q(\text{Tr}(\rho^\alpha))=
\frac{1}{1-\alpha}\textbf{log}_q(1-\frac{d^{\alpha-1}-1}{d^{\alpha-1}}\;S_\alpha(\rho))\;.
\end{eqnarray}
Note that one can also obtain the von Neumann entropy by means of
the $m$-concurrence. Obviously, that requires computation of all
$S_{k}$ from $\alpha=2$ to $\alpha=d$. We choose Renyi's entropy
with $\alpha=2$ and $q=2$. In the following we write $S$ for
$S_2^{\;2}$.

As shown in \cite{HH2,hiesmayr08a}, the advantage of rewriting the
entropies by means of operators is that it is known how to derive
bounds. We present here a way analogous to the method introduced for
the concurrence for bipartite systems in Ref.~\cite{mintert05}. For
that we define a flipped density matrix
\begin{eqnarray}
\widetilde{\rho}_{O_C} := (O_C+O_C^\dagger)\;\rho^*\;(O_C+O_C^\dagger)\;,
\end{eqnarray}
where the conjugation is taken in the computational basis. By
calculating the square root of the eigenvalues of
$\rho\widetilde{\rho}_{O_C}$, which we denote as $\lambda_i^{O_C}$ ,
the bounds for the concurrence are given by:
\begin{eqnarray}\label{bounds}
C_{\textbf{s}\alpha\beta\cdots\omega}(\rho)\geq \max\biggl\lbrace
0,\sum_{O_C}(2\;\max_{\lambda_i^{O_C}}(\{\lambda_i^{O_C}\})-\sum_i\;\lambda_i^{O_C})\biggr\rbrace\;.
\end{eqnarray}

\section{Multipartite entanglement measures}\label{chmeasure}

In this section we propose the two sets of multipartite entanglement
measures. First, we introduce the \emph{separability measure} that
is based on the $\gamma_k$--separability, and second, the
\emph{physical measure} that reveals the ``kind'' of entanglement
between subsystems (bipartite, tripartite,\dots entanglement).

\subsection{Separability measure}\label{separabilitymeasure}

In the following we assume that the total state $\rho$ is pure. For
the generalized multipartite set of entanglement measures there are
a few alternatives, we propose the following
generalization:
\begin{enumerate}
\item[S1a:]
    $E_{tot}(\rho)=\sum_{s=1}^{n}S(\rho_s):=\sum_{\{\alpha_j\}} E_{\{\alpha_j\}}\;>\;0\qquad\forall\;\rho\quad\textrm{with}\;k<n$
    \item[S1b:] $E_{tot}(\rho)\;=\;0\qquad\forall\;\rho\quad\textrm{with}\;k=n$
    \item[S2:] $E_{\{\alpha_j\}}(\rho)\;>\;0\qquad\forall\quad \{\alpha_j\}\in\gamma_k\quad\textrm{and}\quad|\{\alpha_j\}|\;\geq\;2$
    \item[S3:] $E_{\{\alpha_j\}}(\rho)\;=\;0\qquad\forall\quad \{\alpha_j\}\notin\gamma_k\quad\textrm{or}\quad|\{\alpha_j\}|\;=\;1$
    \item[S4:] $E_{\{\alpha_j\}}(\rho^{\otimes n})\;=\;n\;E_{\{\alpha_j\}}(\rho)$ (additivity on copies of the same state)
    \item[S5:] $E_{\{\alpha_j\}}\big(U_{\alpha_1} \otimes U_{\alpha_2} \otimes
\cdots \otimes U_{\alpha_j}\,\rho\, (U_{\alpha_1} \otimes
U_{\alpha_2}\otimes \cdots\otimes
U_{\alpha_j})^\dagger\big)\;=\;E_{\{\alpha_j\}}(\rho)$ (invariance
under local unitary
    operations)
    \item[S6:] $E_{tot}(\rho_1\otimes\rho_2)\;=\;E_{tot}(\rho_1)+E_{tot}(\rho_2)$ (additivity on tensor products of arbitrary states)
\end{enumerate}
With a measure that fulfills all this requirements one obtains the
$\gamma_k$--separability and, moreover, the quantified information
content of a given state.

According to our notation of $\gamma_k$--separability a pure state of three qubits
can be entangled in four different ways,
$\{1|23\};\{12|3\};\{13|2\};\{123\}$, hence we have four different
entanglement measures, which we define in an intuitive way by
\begin{eqnarray}
E_{12}:&=&\{S(\rho_1)+S(\rho_2)\}\cdot\delta[S(\rho_{12}),0]\\
E_{13}:&=&\{S(\rho_1)+S(\rho_3)\}\cdot\delta[S(\rho_{13}),0]\\
E_{23}:&=&\{S(\rho_2)+S(\rho_3)\}\cdot\delta[S(\rho_{23}),0]\\
E_{123}:&=&S(\rho_1)+S(\rho_2)+S(\rho_3)-E_{12}-E_{13}-E_{23}
\end{eqnarray}
with
\begin{eqnarray}
\delta[S(\rho_{\{\alpha_j\}}),0]&=&1\quad\textrm{if}\quad S(\rho_{\{\alpha_j\}})=0\nonumber\\
\delta[S(\rho_{\{\alpha_j\}}),0]&=&0\quad\textrm{if}\quad
S(\rho_{\{\alpha_j\}})>0\;.
\end{eqnarray}

\begin{center}
\begin{tabular}{l}
\parbox[b]{12cm}{\begin{small}\textbf{Instructive example for three qubits:}
For the state
\begin{eqnarray} \label{exghzmix}
|\tilde{\phi}\rangle\;=\;\frac{1}{N}\bigl\lbrace p\,|GHZ\rangle+(1-p)\,|\phi^+
\rangle\otimes\{\cos\alpha|0\rangle+\sin\alpha|1\rangle\}\}
\end{eqnarray}
with
$|GHZ\rangle=\frac{1}{\sqrt{2}}\{|000\rangle+|111\rangle\bigr\rbrace$
we expect for $p=0$ that it is $2$--separable with
$\gamma_2=\{12|3\}$ (in detail $E_{12}=2, E_{13}=E_{23}=E_{123}=0$)
and for $p=1$ it is $1$--separable (=fully entangled)
($E_{123}=1,E_{12}=E_{13}=E_{23}=0 $). For values of $p\in\{0,1\}$
it depends on $\alpha$. With the separability measure this can be
easily calculated:
\begin{itemize}\item[(a)] If we want $E_{12}$ to vanish,
then $S(\rho_{12})$ has to be zero, this can be obtained for $p\leq 0.58$
and $\cos(\pm\alpha)=\frac{p^2-p\pm\sqrt{p^4-6 p^3+11 p^2-8 p+2}}{2
\left(p^2-2 p+1\right)}$. The state is fully separable, except for $p=0.5$ where $S(\rho_{12})=0$, but
also $S(\rho_1), S(\rho_2), S(\rho_3)=0$.
\item[(b)] No $\alpha$ and $p$ exist such that $S(\rho_{13})$ or $S(\rho_{23})$ vanish, thus $E_{13}=E_{23}=0$.
\item[(c)] As $E_{13}$ and $E_{23}$ are always zero, the state is
either $\gamma_1=\{123\}$--separable ($E_{123}>0$) or
$\gamma_2=\{12|3\}$--separable ($E_{12}>0$); except for $p=0.5$ and
the above $\alpha$ then the state is fully separable
$\gamma_3=\{1|2|3\}$.
\item[(d)] If the GHZ state is interchanged with a W state the state
is for all $p$ and $\alpha$ $3$--separable except for $p=0$, where it
is clearly $\gamma_2$--separable.
\end{itemize}
\end{small}}
\end{tabular}
\end{center}

The separability measure can be generalized for multipartite qudit systems as
\begin{eqnarray}
E_{\{\alpha_j\}}:=\sum_{s\in\{\alpha_j\}}\biggl(S(\rho_s)-
\sum_{\{\beta_j\}\subset\{\alpha_j\}}E_{\{\beta_j\}}\biggr)\cdot\delta[S(\rho_{\{\alpha_j\}}),0] \,,
\end{eqnarray}
where
\begin{eqnarray}
\sum_{\{\alpha_j\}}E_{\{\alpha_j\}}=E_{tot}(\rho) \,.
\end{eqnarray}
This is an important feature as any violation of this necessary
requirement would imply either neglecting or over-quantifying of the
information content. As is proven in the Appendix~\ref{ProofSep}
this proposed set of measures meets all requirements S1-S6.

The separability measure provides a set of entanglement measures for pure states. In
principle it can be extended to mixed states using the convex roof method,
\begin{eqnarray}
E_{\{\alpha_j\}}(\rho)\;:=\;\inf\sum_i p_i\;
E_{\{\alpha_j\}}(|\psi_i\rangle\langle\psi_i|)\;.
\end{eqnarray}
Since there still is no method to calculate the convex roof for
arbitrary states, the proposed measure is computable only for pure
states or mixed states for which the bound of the $m$--concurrences
are exact or for states where we can know if all involved entropies
vanish or not. Clearly, if one cannot execute the defined Kronecker
$\delta$'s exactly, the computation of the measure may fail.
Moreover, when applying the separability measure for mixed states,
one does in general not obtain e.g. whether the state at all is
entangled as the the following example shows:

\begin{center}
\begin{tabular}{l}
\parbox[b]{12cm}{\begin{small}\textbf{Instructive example for even number of qubits:}
Consider the generalized Smolin state
$\rho_{Smolin}=\frac{1}{2^n}\left(\mathbbm{1}+\sum c_i\;
\sigma_i^{\otimes n}\right)$, where $n$ is a even number. This state
is mixed (except for $n=2$ and $|\vec{c}|=3$) and the states of all subsystems are maximally mixed. Therefore all $E_{12\dots j}$ with
$j<n$ derive to zero. Thus for the Smolin state only $E_{12\dots n}$
can be nonzero. To show for which parameters the state is entangled,
we need another measure which we introduce in the next section and
apply in Sect.~\ref{smolin} to the Smolin state.
%
%
\end{small}}
\end{tabular}
\end{center}

\subsection{Physical measure}

As we have mentioned, there are many different aspects of
multipartite entanglement one might be interested in. The separability
measure quantifies the total information content in entanglement and
yields an answer to the $\gamma_k$--separability in a multipartite
state. From a physical point of view, however, we can also present
another approach to quantify multipartite entanglement. The basic
motivation is to reveal structures of quantum states that go beyond
separability. Take for instance the instructive
example of a $W$ state $|W\ra$ and the Greenberger-Horne-Zeilinger state $|GHZ\ra$ in the three qubit system, where $|W\ra = 1/\sqrt{3} \left( |001\ra + |010\ra + |100\ra \right)$ and $|GHZ\ra$ as defined in Eq.~\eqref{exghzmix}. Both
are completely inseparable and thus it is not possible to distinguish them by the separability measure. From a physical point of view the most obvious difference between these two states is the following: Ignoring an
arbitrary subsystem will, in case of the $GHZ$ state, yield a mixed
separable state, whereas in case of the $W$ state, will yield a
mixed entangled state. Any set of entanglement measures that is
designed to capture this difference will need a modification in
requirements (S2) and (S3) and for mixed density matrices we need the additional requirements (P4) and (P5) as defined below. Thus the properties we propose are the following:
\begin{center}
\begin{enumerate}
\item[S1a:]
    $\mathcal{E}_{tot}(\rho)=\sum_{s=1}^{n}S(\rho_s)\;>\;0\qquad\forall\;\rho\quad\textrm{with}\;k<n$
    \item[S1b:] $\mathcal{E}_{tot}(\rho)\;=\;0\qquad\forall\;\rho\quad\textrm{with}\;k=n$
    \item[P2:] $\mathcal{E}_{\{\alpha_j\}}(\rho)\;\geq\;0\quad\forall\quad \{\alpha_j\}\subseteq\{\beta_i\}\in\gamma_k\quad\textrm{and}\quad |\{\alpha_j\}|\geq 2$
    \item[P3:] $\mathcal{E}_{\{\alpha_j\}}(\rho)\;=\;0\quad\forall\quad\{\alpha_j\}\supset\{\beta_i\}\in\gamma_k \quad\textrm{or}\quad |\{\alpha_j\}|=1$
\item[P4:] $\mathcal{E}_{\{\alpha_j\}}(\lambda\; \rho_1 + (1-\lambda)\;\rho_2)\;\leq\;\lambda\; \mathcal{E}_{\{\alpha_j\}}(\rho_1)
+ (1-\lambda)\;\mathcal{E}_{\{\alpha_j\}}(\rho_2)\qquad$ (convexity)
\item[P5:] $\sum_i \tr \left( V_i \rho V_i^\dag \right) \mathcal{E}_{tot}
\left( \frac{V_i \rho V_i^\dag}{\tr \left( V_i \rho V_i^\dag \right)} \right) \leq \mathcal{E}_{tot}(\rho)$ (non-increasing on average under LOCC),\\
       where $V_i$ is a separable operator, i.e. of the local form $V_i := V_i^1 \otimes V_i^2 \otimes \ldots \otimes V_i^n$\,.
\end{enumerate}
 \end{center}
Of course capturing this essential difference needs computation of
entanglement of all possible subsystems, which are in general mixed.
Therefore we propose a set of measures which contain a convex roof
extension already for the subsystems of pure multipartite states.

For that let us first define the following useful quantity for any
density matrix $\rho=\sum_i p_i |\psi_i\rangle\langle\psi_i|$:
\begin{eqnarray}
P(\rho):=\inf_{p_i,\psi_i,\gamma_k} \sum_i p_i\; (\sum_s
S(\text{Tr}_{\neg s}|\psi_i\rangle\langle\psi_i|))\;.
\end{eqnarray}
Here the sum over all subsystems $s$ is taken over the entropy of
pure states $\psi_i$, thus is the ``correct'' entanglement content
of this certain state $\psi_i$. Note that we take the infimum also
over $\gamma_k$. All examples we have considered show that the
infimum over $\gamma_k$ has not separably to be claimed, because the
infimum over all decompositions was always achieved in the correct
$\gamma_k$-separable decomposition. Moreover, for physical reasons
it would be surprising if this was not the case. It would mean that
there exist e.g. a partially separable state for which the infimum
of the subsystem's entropies was realized for a completely
inseparable decomposition. However, we were yet not able to prove that rigorously and
therefore we have to conjecture that for any state with the
following $\gamma_k$ separability
\begin{eqnarray}
\gamma_k:=\{\{\beta_1\}|\{\beta_2\}|\cdots|\{\beta_k\}\}
\end{eqnarray}
the equality
\begin{eqnarray}
P(\rho_{\gamma_k})=P(\rho_{\{\beta_1\}})+P(\rho_{\{\beta_2\}})+(\cdots)+P(\rho_{\{\beta_k\}})\;.
\end{eqnarray}
holds. Note that in this way we trivially obtain the additivity property we proposed for
the physical measure (see Appendix).

For convenience, we start to define the set of measures for four
particles by
\begin{eqnarray}
\textrm{two--particle
entanglement:}&&\mathcal{E}_{12}=P(\rho_{12})\;,\quad\mathcal{E}_{13}=P(\rho_{13})\;,\\
&&
\mathcal{E}_{14}=P(\rho_{14})\;,\quad\mathcal{E}_{23}=P(\rho_{23})\;,\\
&&\mathcal{E}_{24}=P(\rho_{23})\;,\quad\mathcal{E}_{34}=P(\rho_{34})\;,
\\
\textrm{three--particle
entanglement:}&&\mathcal{E}_{123}=\max\bigl[0,P(\rho_{123})-\mathcal{E}_{12}-\mathcal{E}_{13}-\mathcal{E}_{23}\bigr]\;,\\
&&
\mathcal{E}_{124}=\max\bigl[0,P(\rho_{124})-\mathcal{E}_{12}-\mathcal{E}_{14}-\mathcal{E}_{24}\bigr]\;,\\
&&
\mathcal{E}_{134}=\max\bigl[0,P(\rho_{134})-\mathcal{E}_{13}-\mathcal{E}_{14}-\mathcal{E}_{34}\bigr]\;,\\
&&
\mathcal{E}_{234}=\max\bigl[0,P(\rho_{234})-\mathcal{E}_{23}-\mathcal{E}_{24}-\mathcal{E}_{34}\bigr]\;,\\
\textrm{four--particle entanglement:}&&
\mathcal{E}_{1234}=\max\bigl[0,P(\rho_{1234})-\mathcal{E}_{123}-\mathcal{E}_{124}-\mathcal{E}_{134}-\mathcal{E}_{234}\\
&&\hphantom{\mathcal{E}_{1234}=}-\mathcal{E}_{12}-\mathcal{E}_{13}-\mathcal{E}_{14}-\mathcal{E}_{23}-\mathcal{E}_{24}-\mathcal{E}_{34}\bigr]\end{eqnarray}

\begin{center}
\begin{tabular}{l}
\parbox[b]{12cm}{\begin{small}\textbf{Instructive example for $4$ qubits:}
Consider the mixture of the $GHZ$ state and a pair of $EPR$-states state:
\begin{eqnarray}\label{ghzeprstate} \rho&=&\cos^2(\alpha)\;|GHZ\rangle\langle
GHZ|+\sin^2(\alpha)\;|\Phi^+\rangle\langle \Phi^+|\otimes|\Phi^+\rangle\langle \Phi^+|\nonumber\\\end{eqnarray} with
\begin{eqnarray}
|GHZ\rangle&=&\frac{1}{\sqrt{2}}(|0000\rangle+|1111\rangle)\nonumber\\
|\Phi^+\rangle&=&\frac{1}{\sqrt{2}}(|00\rangle+|11\rangle)\;.\end{eqnarray}
The set of the physical measure is visualized in Fig.~\ref{eprghz}.
\end{small}}
\end{tabular}
\end{center}

\begin{figure}
\includegraphics[keepaspectratio=true]{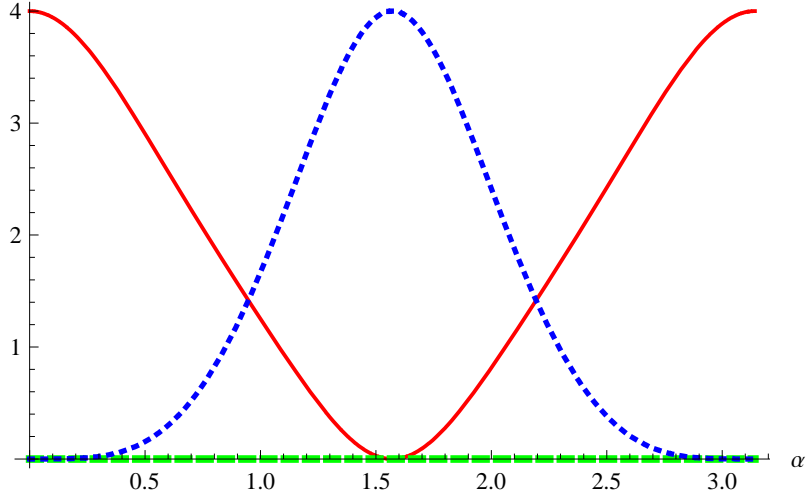}
\caption{(Color online) The graphes show the set of the physical
measure of the mixture of the $GHZ$ state and the $EPR\otimes EPR$
state, Eq.~(\ref{ghzeprstate}). The solid (red) curve shows the
four--partite entanglement $\mathcal{E}_4=\mathcal{E}_{1234}$, the
dashed (green) curve shows the three--partite entanglement
$\mathcal{E}_3=\mathcal{E}_{123}+\mathcal{E}_{124}+\mathcal{E}_{134}+\mathcal{E}_{234}$
and the dotted (blue) curve shows the two--partite entanglement
$\mathcal{E}_2=\mathcal{E}_{12}+\mathcal{E}_{13}+\mathcal{E}_{14}+\mathcal{E}_{23}+\mathcal{E}_{24}+\mathcal{E}_{34}$
in dependence of $\alpha$. The amount of the total entanglement is
for the $GHZ$ state and the $EPR\otimes EPR$ state $4$, however, in
the first case it due to four--partite entanglement whereas in the
other case the bipartite entanglement maximizes. The separability
measure reveals that the state is $\gamma_1=\{1234\}$--separable
($E_{1234}=4$, all others zero) except for $\alpha=\frac{\pi}{2}$
then the state is $\gamma_2=\{12|34\}$--separable
($E_{tot}=4,E_{12}=E_{34}=2$ all others zero).}\label{eprghz}
\end{figure}

The generalization for any multipartite qudit system is
straight forward:
\begin{eqnarray}
\mathcal{E}_{\{\alpha_j\}}=\max\bigl[0,P(\rho_{\{\alpha_j\}})-\sum_{\{\beta_j\}
\subset\{\alpha_j\}}\mathcal{E}_{\{\beta_j\}}\bigr]\;.
\end{eqnarray}
Note that in case of the physical measure
$\sum_{\{\alpha_j\}}\mathcal{E}_{\{\alpha_j\}}=\mathcal{E}_{tot}(\rho)$
is no longer a requirement. Indeed there exist states which even
violate this condition. This is due to the fact that the physical
measure quantifies the entanglement of subsystems of a larger
systems with respect to possible applications or distillation. In
case of overlapping indices of subsystems, e.g. $\rho_{123}$ and
$\rho_{124}$ the possibility arises that both share the same
entanglement, e.g. in subsystem $1$ and $2$, and thus in sum
overquantify the actual total entanglement. However, there is no
contradiction to possible experiments as one would have to decide,
which subsystems to use, e.g. $\rho_{123}$ or $\rho_{124}$, as their
entanglement properties cannot be exploited simultaneously.

With the help of the $m$--concurrence, Eq.~(\ref{mconcurrence}),
bounds for every $P(\rho)$ can be computed, see Eq.~(\ref{bounds}),
and thus for the whole set of entanglement measures. In the next
section we give further examples and explicit formulae.

\section{Further instructive examples}

In this section we apply our two sets of entanglement measures to certain classes
of states and show explicitly how to derive the desired quantities.

\subsection{Two-qubit states}

In the case of pure bipartite qubit states obviously both measures
coincide:
\begin{eqnarray}
E(\rho_{12})\;=\;E_{12}\;=\; \mathcal{E}_{12}&=&S(\rho_1)+S(\rho_2)\nonumber\\
&=&-\log_2(\tr(\rho_1^2))-\log_2(\tr(\rho_2^2))\\
&=&-\;\log_2(1-\frac{1}{2}\textbf{C}_{12}^2)-\;\log_2(1-\frac{1}{2}\textbf{C}_{12}^2)\nonumber\\
&=&-2\;\log_2(1-\frac{1}{2}\textbf{C}_{12}^2)
\end{eqnarray}
where the concurrence $\textbf{C}_{12}$  is twice the Hill--Wootters
concurrence \cite{hill97}. hgbz76 For mixed states $\rho_{12}=\sum_i
p_i |\psi_i\ra \la \psi_i|$, we obtain the physical measure by
\begin{eqnarray}
 \mathcal{E}_{12}(\rho_{12})&=&P(\rho_{12})\;=\;\inf_{p_i,\psi_i}\sum p_i\; \{S(\tr_2(\rho_i))+S(\tr_1(\rho_i))\}\nonumber\\
 &=&2\;\inf_{p_i,\psi_i}\sum p_i\; S(\tr_2(\rho_i))
\;=\;-2\;\inf_{p_i,\psi_i}\sum p_i
\log_2(\tr\{(\tr_2(\rho_i))^2\})\nonumber\\
&=&-2\inf_{p_i,\psi_i}\sum p_i
\log_2(1-\frac{1}{2}\textbf{C}_{12}^2(\psi_i))\nonumber\\
&\geq&-2\;\inf_{p_i,\psi_i} \log_2(1-\frac{1}{2}\sum
p_i\textbf{C}_{12}^2(\psi_i))\;=\;-2\;
\log_2(1-\frac{1}{2}\textbf{C}_{12}^2(\rho_{12})) \,,
\end{eqnarray}
where the concurrence can be derived operationally via
\begin{eqnarray}
\textbf{C}_{12}(\rho_{12})=\max\biggl\lbrace 0,2
\max_{\lambda_i^{O_C}}(\{\lambda_i^{O_C}\})-\sum_i\;\lambda_i^{O_C}\biggr\rbrace\
\,,
\end{eqnarray}
where the $\lambda_i^{O_C}$ are the square roots of the eigenvalues
of $\rho_{12}\;\tilde{\rho}_{12}$ and
$\tilde\rho_{12}=(O_C+O_C^\dagger)\;\rho_{12}^*\;(O_C+O_C^\dagger)$
with $O_C+O_C^\dagger=\sigma_y\otimes\sigma_y$. For bipartite qubits
it is known that there always exists a decomposition such that all
concurrences of the pure states $|\psi_i\ra$ are equal
\cite{hill97}, therefore the inequality is in fact an equality and
the bounds are also known to be exact.

\subsection{Two-qutrit states}

In the case of qutrits the linear entropies can be written by only
six different operators which are all possible tensor products of
the three symmetric Gell-Mann matrices
$\sigma^{(i)}=|j\rangle\langle k|+|k\rangle\langle j|$ with $0\leq
j<k\leq 2$:
\begin{eqnarray}
S(\rho_1)&=&-\log_2(1-\frac{1}{2}(\sum_{ij}
\textbf{C}_{\textbf{1}2}^{\sigma^{(i)}\otimes\sigma^{(j)}}))\nonumber\\
S(\rho_2)&=&-\log_2(1-\frac{1}{2}(\sum_{ij}
\textbf{C}_{1\textbf{2}}^{\sigma^{(i)}\otimes\sigma^{(j)}}))
\end{eqnarray}
Lower bounds on $\mathcal{E}_{12}(\rho_{12})=P(\rho_{12})$ are then
obtained by calculating the squared eigenvalues of all operators
$O^{\sigma^{(i)}\otimes\sigma^{(j)}}$ and adding them.

Consider the class of qutrit states which are composed of any two
generalized Bell states denoted by $P_{00}, P_{01}$ and the totally
mixed state (for an review on the geometry of that class of states
see \cite{BHN1,BHN2,BHN3})
\begin{eqnarray}\label{qutritsPP}
\rho(\alpha,\beta)&=&\frac{1-\alpha-\beta}{9}\mathbbm{1}+\alpha
P_{00}+\beta P_{01}\;.
\end{eqnarray}
Here $P_{k,l}=|\Omega_{k,l}\rangle\langle\Omega_{k,l}|$ are obtained
by choosing one maximally entangled state, e.g. $\Omega _{0 0} =
\frac{1}{\sqrt{d}} \sum_{s=0}^{d-1} |s\rangle \otimes |s\rangle$,
and applying Wely-Operators $W_{k,l}|s\rangle =
w^{k(s-l)}|s-l\rangle$ onto one subsystem, i.e.
$\Omega_{k,l}=W_{k,l}\otimes\mathbbm{1}_d\; \Omega_{0,0}$.

\begin{figure}
\begin{center}
(a)\includegraphics[width=7cm,
keepaspectratio=true]{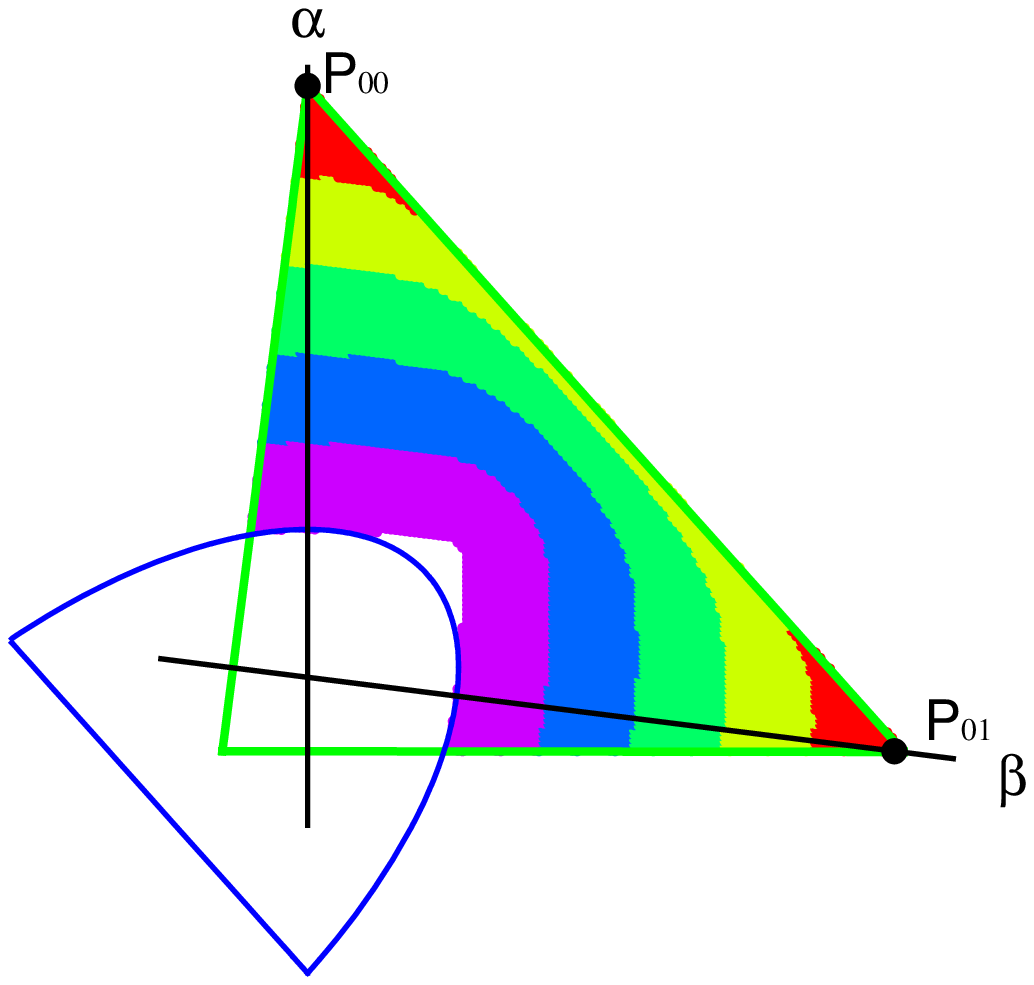}
(b)\includegraphics[width=6.5cm,
keepaspectratio=true]{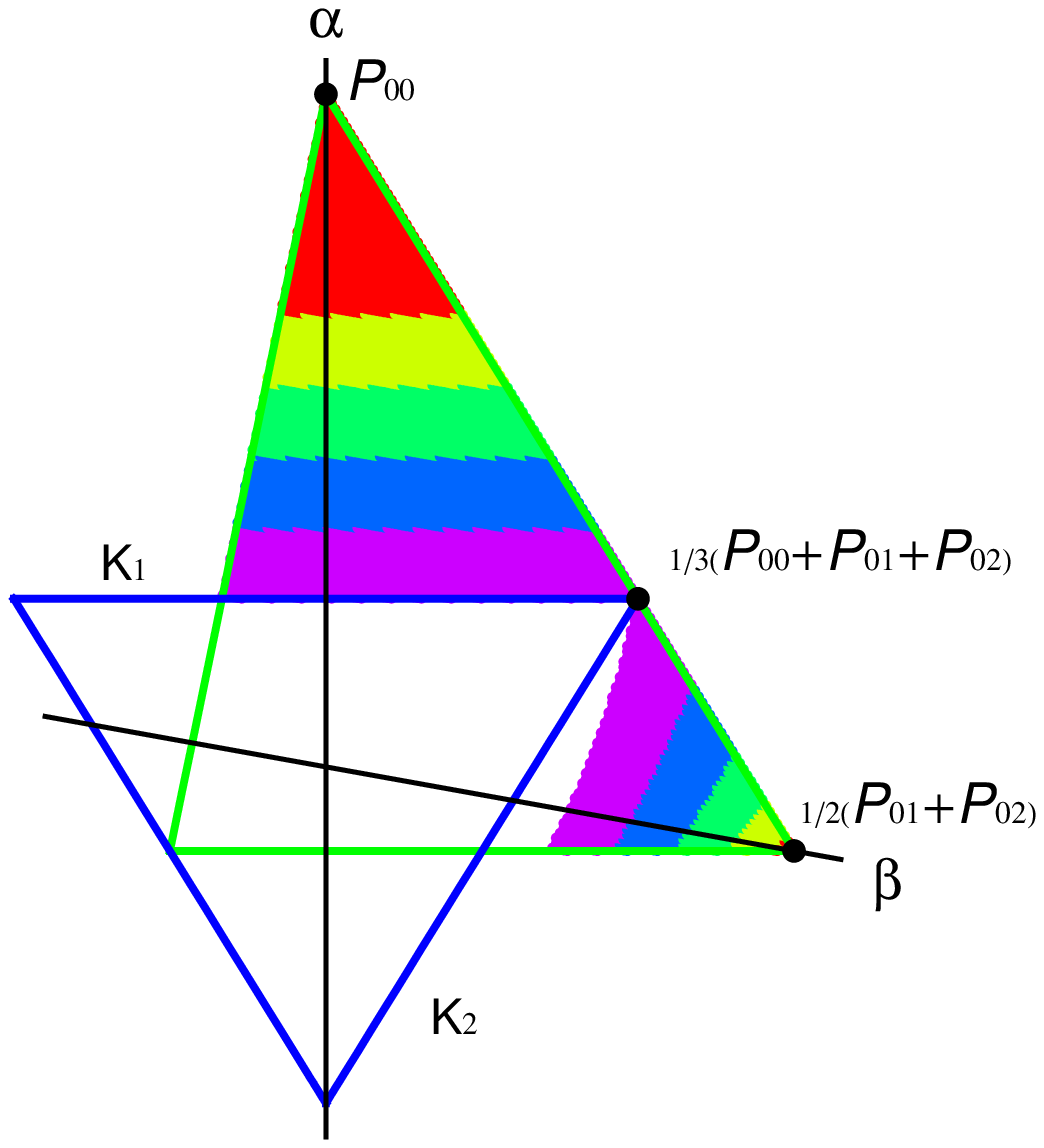}\caption{(Color online) Here
two slices through the class of ``line'' states \cite{BHN3},
Eq.~(\ref{qutritsPP}) and Eq.~(\ref{qutritsPPP}), are shown. The
green triangle visualizes the parameter space for which the states
are positive, the blue triangle/ellipse the parameter space for
which the states are positive under partial transpose (PPT). The
colored areas denote the regions where the bounds on the physical
measure are nonzero (red: $1\geq \textbf{C}> 0.8$; yellow: $0.8\geq
\textbf{C}> 0.6$; green: $0.6\geq \textbf{C}> 0.4$; blue: $0.4\geq
\textbf{C}> 0.2$; purple: $0.2\geq \textbf{C}> 0$). Note that not
all states negative under partial transpose are detected. In
Fig.~(a) for $\alpha<0$ or $\beta<0$ the bound is equivalent to the
boundary by PPT, however, as was shown in Ref.~\cite{BHN3} a small
region of bound entangled states exist in this case. Only for the
class of states visualized in Fig.~(b) for $\alpha\geq\frac{1}{4}$
the bounds are tight.} \label{qutrits}
\end{center}
\end{figure}

The result of the physical measure is visualized in
Fig.~\ref{qutrits}~(a). If $\alpha$ or $\beta$ is zero, then we
obtain the famous isotropic states, for which we now that
concurrence increases linearly with $\alpha(\beta)$. If $\alpha$ and
$\beta$ are both positive, we observe that not all states negative
under partial transpose ($NPT$) are detected, thus the bounds are
not exact. If either $\alpha$ or $\beta$ is zero, the derived bounds
detect all $NPT$ states, however, as shown in \cite{BHN1} in this
case bound entangled states exist, therefore the bounds are not
exact. Recently, by using bounds obtained by an operator acting
globally on two copies of a state, these bound entangled states
could be detected \cite{Bae09}, however, in the region
$\alpha,\beta>0$ the bounds did not detect all NPT states either,
but they were tighter than the bounds introduced in this work.

Let us consider the class of states
\begin{eqnarray}\label{qutritsPPP}
\rho(\alpha,\beta)&=&\frac{1-\alpha-\beta}{9}\mathbbm{1}+\alpha
P_{00}+\frac{\beta}{2}(P_{01}+P_{02})\;,
\end{eqnarray}
where the third Bell states is obtained by applying the same Weyl
operator, which transforms $P_{00}$ to $P_{01}$, to $P_{01}$. This
class of states are visualized in Fig.~\ref{qutrits}~(b). Here more
symmetries are involved, therefore no bound entangled states can be
found and the optimal entanglement witnesses, ${\cal K}_1,{\cal
K}_2$, correspond to lines. It turns out that the bounds are only
exact for $\alpha>\frac{1}{4}$. The requirements for the bounds to
be tight is left for further investigation.

\subsection{Three-qubit states}

Let us consider the most general tripartite pure qubit state,
\begin{eqnarray}
|\psi\rangle=a|000\rangle+b|001\rangle+c|010\rangle+d|011\rangle+e|100\rangle+f|101\rangle+g|110\rangle+h|111\rangle\;.
\end{eqnarray}
The linear entropies of all three subsystems can be rewritten in
terms of $m$--concurrences
\begin{eqnarray}
S(\rho_1)&=&-\textbf{log}_2(1-\frac{1}{2}(\textbf{C}_{12}^2+\textbf{C}_{13}^2+\textbf{C}_{\textbf{1}23}^2)),\\
S(\rho_2)&=&-\textbf{log}_2(1-\frac{1}{2}(\textbf{C}_{12}^2+\textbf{C}_{23}^2+\textbf{C}_{1\textbf{2}3}^2)),\\
S(\rho_3)&=&-\textbf{log}_2(1-\frac{1}{2}(\textbf{C}_{23}^2+\textbf{C}_{13}^2+\textbf{C}_{12\textbf{3}}^2))\;.
\end{eqnarray}
Also the entropies of the partially reduced subsystems can be
rewritten into the $m$--concurrences
\begin{eqnarray}
S(\rho_{12})&=&-\textbf{log}_2(1-\frac{1}{2}(\textbf{C}_{23}^2+\textbf{C}_{13}^2+\textbf{C}_{12\textbf{3}}^2)),\\
S(\rho_{13})&=&-\textbf{log}_2(1-\frac{1}{2}(\textbf{C}_{12}^2+\textbf{C}_{23}^2+\textbf{C}_{1\textbf{2}3}^2)),\\
S(\rho_{23})&=&-\textbf{log}_2(1-\frac{1}{2}(\textbf{C}_{12}^2+\textbf{C}_{13}^2+\textbf{C}_{12\textbf{3}}^2))\;.
\end{eqnarray}
Note that for pure tripartite qubit systems there is an intuitive
relation between the entropies of the subsystems, $S(\textbf{Tr}_k
|\psi\rangle\langle\psi|)=S(\textbf{Tr}_{\neg k}
|\psi\rangle\langle\psi|)$. For explicit examples see
Ref.~\cite{HH2}.

\subsection{The generalized Smolin states}\label{smolin}

As shown in Ref.~\cite{hiesmayr08a} for the generalized Smolin state
$\rho_{Smolin}=\frac{1}{2^n}\left(\mathbbm{1}+\sum
c_i\;\sigma_i^{\otimes n}\right)$ ($n$ even) only the $n$--flip
concurrence $\textbf{C}_{12\dots n}$ is nonzero and the bounds turn
out to be tight. In detail one obtains the $n$--partite entanglement
($X=1$ for $n\geq4$ and $X=2$ for $n=2$)
\begin{eqnarray}
E_{12\dots n}&=&\mathcal{E}_{12\dots
n}\;=-4\log_2\biggl\lbrace\frac{1}{4}\;X\max\big[0,\;\frac{1}{2}\max\bigl\lbrace-1+\vec{c}\cdot\vec{n}^{\,(1)},
-1+\vec{c}\cdot\vec{n}^{\,(2)},
-1+\vec{c}\cdot\vec{n}^{\,(3)},\nonumber\\
&&\qquad\qquad -1+\vec{c}\cdot\vec{n}^{\,(4)}\,\bigr\rbrace\biggr]^2\biggr\rbrace
\end{eqnarray}
with
\begin{eqnarray}
\{\vec{n}^{\,(0)},\vec{n}^{\,(1)},\vec{n}^{\,(2)},\vec{n}^{\,(3)}\}&=&
\{\left(\begin{array}{c}-1\\-1\\-1\\\end{array}\right),
\left(\begin{array}{c}+1\\+1\\-1\\\end{array}\right),
\left(\begin{array}{c}-1\\+1\\+1\\\end{array}\right),
\left(\begin{array}{c}+1\\-1\\+1\\\end{array}\right)\}\;.
 \nonumber\end{eqnarray}
 Therefore, the state is fully or $n$--separable for
 $-1+\vec{c}\cdot\vec{n}^{\,(i)}\leq 0$ for all $i$ and $n$--partite entangled else.

\section{Conclusion}

In this paper we first extend the notion of $k$--separability to the
$\gamma_k$--separability which includes the knowledge which
subsystems are inseparable joint. We have pointed out that for mixed
states the extension is not straightforward as an ambiguity could
arise, however, we could overcome this problem by an appropriate
definition, i.e. by a maximization over $k$. Moreover, this novel
$\gamma_k$--separability concept shows also the desired convexity
property for mixtures of different $\gamma_k$ states, as visualized
in Fig.~\ref{gammak}.

Based on this extended concept we could define two different sets of
entanglement measures, the first one reveals the $\gamma_k$
separability property, the second one reveals the structural,
physical properties, e.g. the kind of entanglement.

Both measures are based on the convex roof extension which in
general cannot be computed. We use the method of Ref.~\cite{HH2}
based on the observation that any entropy can be rewritten by
$m$--flip concurrences, i.e. in an operator form. This includes any
qudit system. Therefore bounds on the set of measures can be
obtained and we show their usefulness in several examples.


For certain applications, such as quantum cryptography scenarios,
one is rather interested in the structure of entanglement. For that
we have defined a set of measures revealing the two--partite
(bipartite), three--partite (tripartite), \dots , $n$--partite
entanglement, which we denote as a ``physical'' measure. It captures
for example the different entanglement features of e.g. the $GHZ$
states, the $W$ states or of the $EPR\otimes EPR$ states.

In the last section we gave more instructive examples with explicit
formulae to compute lower bounds of the entanglement measures. We
show cases where the bounds are surprisingly tight and cases were
they are not. Further investigations have to be performed in order
to understand in which cases the bounds are equivalent to the
infimum of the convex roof.

In summary, we have pointed out that all entanglement features in
multipartite systems cannot be revealed by a single set of measures.
We defined two sets of measures for multipartite qudit systems and
demonstrated its usefulness and computability. Herewith we believe
one may find novel application exploiting the entanglement of
multipartite systems, which is ---as this work shows--- at least
mathematically considerably different to bipartite qubit
entanglement.

\section{Acknowledgements}

We would like to thank Heide Narnhofer for enlightening discussions
on the topic. P. K. would like to acknowledge financial support by
FWF project CoQuS No. W1210-N16 of the Austrian Science Foundation.

\section{Appendix}

In this appendix we give the proofs that the proposed set of
separability measures and the set of the physical measures fulfill
the proposed requirements.

\subsection{Proofs for the separability measure}\label{ProofSep}

Here we prove the proposed requirements S2-S6 for the separability
measure, S1 is a definition.\\
\\\textbf{Property:} $$S2:\quad E_{\{\alpha_j\}}(\rho)\;>\;0\qquad\forall\quad
\{\alpha_j\}\in\gamma_k\quad\textrm{and}\quad|\{\alpha_j\}|\;\geq\;2
$$ with $E_{\{\alpha_j\}}:=\sum_{s\in\{\alpha_j\}}\biggl(S(\rho_s)-
\sum_{\{\beta_j\}\subset\{\alpha_j\}}E_{\{\beta_j\}}\biggr)\cdot\delta[S(\rho_{\{\alpha_j\}}),0]\;.$
\begin{proof}
Consider a $n$-partite pure state $\rho$ with the separability
property
$\gamma_k:=\{\{\varepsilon_1\}|\{\varepsilon_2\}|\cdots|\{\varepsilon_k\}\}$.
Now if $\{\alpha_j\}\in\gamma_k$ then it follows that $
\delta[S(\rho_{\{\alpha_j\}}),0]=1$ and consequently
$\forall\quad\{\beta_j\}\subset\{\alpha_j\}$  and
$\forall\quad\{\beta_j\}\supset\{\alpha_j\}$ is $
\delta[S(\rho_{\{\beta_j\}}),0]=0$. Therefore
$E_{\{\alpha_j\}}=\sum_{s\in{\alpha_j}} S(\rho_s)>0$ as required.
\end{proof}
\noindent\textbf{Property:}
$$ S3:\quad
E_{\{\alpha_j\}}(\rho)\;=\;0\qquad\forall\quad
\{\alpha_j\}\notin\gamma_k\quad\textrm{or}\quad|\{\alpha_j\}|\;=\;1$$
\begin{proof}
Again consider a $n$-partite pure state $\rho$ with the separability
property
$\gamma_k:=\{\{\varepsilon_1\}|\{\varepsilon_2\}|\cdots|\{\varepsilon_k\}\}$.
Now if $\{\alpha_j\}\notin\gamma_k$ then it follows that $
\delta[S(\rho_{\{\alpha_j\}}),0]=0$ and therefore
$E_{\{\alpha_j\}}=0$. If $\{\alpha_j\}\in\gamma_k$ but
$|\{\alpha_j\}|\;=\;1$ then $
E_{\{\alpha_j\}}=S(\rho_{\alpha_j})\;\delta[S(\rho_{\alpha_j}),0]=0\cdot
1=0$.
\end{proof}

\noindent\textbf{Property:}
$$S4: E_{\{\alpha_j\}}(\rho^{\otimes
n})\;=\;n\;E_{\{\alpha_j\}}(\rho)$$

\begin{proof}
Again consider a $n$-partite pure state $\rho$ with the following
separability property
$\gamma_k:=\{\{\varepsilon_1\}|\{\varepsilon_2\}|\cdots|\{\varepsilon_k\}\}$.
$\rho^{\otimes n}$ must have the same $\gamma_k$ property. Thus any
nonzero $E_{\{\alpha_j\}}$ will be of the form $
E_{\{\alpha_j\}}=\sum_{s=\alpha_1}^{\alpha_j}S(\rho_s^{\otimes n})$.
Hence it sufficient to prove that
\begin{eqnarray}
S(\rho_s^{\otimes n}) = n\,S(\rho_s)
\end{eqnarray}
where $S(\rho_s):=\log_2(\text{Tr}(\rho_s^2))$. This is the case as
$ \textnormal{Tr}((\rho_s^{\otimes
n})^2)=(\textnormal{Tr}((\rho_s)^2))^n$.
\end{proof}

\noindent\textbf{Property:}
$$S5:
E_{\{\alpha_j\}}\big(U_{\alpha_1} \otimes U_{\alpha_2} \otimes
\cdots \otimes U_{\alpha_j}\,\rho\, (U_{\alpha_1} \otimes
U_{\alpha_2}\otimes \cdots\otimes
U_{\alpha_j})^\dagger\big)\;=\;E_{\{\alpha_j\}}(\rho)\;.
$$
\begin{proof}
This proof is trivial as every underlying property
$S(\rho):=\log_2(\textnormal{Tr}(\rho^2))$ is clearly invariant
under local unitary transformations, i.e.
$$
\textnormal{Tr}(U\rho \underbrace{U^\dagger U}_{\id}\rho
U^\dagger))=\textnormal{Tr}(\rho^2 \underbrace{U^\dagger
U}_{\id})=\textnormal{Tr}(\rho^2)$$
\end{proof}

\noindent\textbf{Property:}
$$ S6:
E_{tot}(\rho_1\otimes\rho_2)\;=\;E_{tot}(\rho_1)+E_{tot}(\rho_2)\quad
\textrm{with}\quad \sum_{\{\alpha_j\}}
E_{\{\alpha_j\}}=E_{tot}=\sum_{s=1}^{n}S(\rho_s)\;.$$

\begin{proof}
Consider now a $n_1$--partite pure state $\rho_1$ with the
separability property $
\gamma_{k_1}:=\{\{\varepsilon_1\}|\{\varepsilon_2\}|\cdots|\{\varepsilon_{k_1}\}\}$
and another $n_2$--partite pure state $\rho_2$ with the separability
property $
\gamma_{k_2}:=\{\{\kappa_1\}|\{\kappa_2\}|\cdots|\{\kappa_{k_2}\}\}$.
The tensor product of those two states have the separability
property
$$\gamma_{k_3}:=\{\{\varepsilon_1\}|\{\varepsilon_2\}|\cdots|\{\varepsilon_{k_1}\}|\{\kappa_1\}
|\{\kappa_2\}|\cdots|\{\kappa_{k_2}\}\}\;.$$ In this notation the
counting of the subsystems of the second system starts with $n_1+1$.
For every $n$-partite pure state the total entanglement is
$$E_{tot}(\rho)=\sum_{s=1}^{n}S(\rho_s)$$
and hence
$$
E_{tot}(\rho_1\otimes\rho_2)=\sum_{s=1}^{n_1+n_2}S(\rho_s)=\underbrace{\sum_{s=1}^{n_1}S(\rho_s)}_{E_{tot}(\rho_1)}+\underbrace{\sum_{s=n_1+1}^{n_2}S(\rho_s)}_{E_{tot}(\rho_2)}
\;.$$
\end{proof}

\subsection{Proofs for the physical measure}\label{ProofPhys}
\noindent\textbf{Property:}
\begin{enumerate}
    \item[P1:] $\mathcal{E}_{\{\alpha_j\}}(\rho)\;\geq\;0\qquad\forall\quad \{\alpha_j\}\subseteq\gamma_k\wedge|\{\alpha_j\}|\;\geq\;2$
    \item[P2:] $\mathcal{E}_{\{\alpha_j\}}(\rho)\;=\;0\qquad\forall\quad \{\alpha_j\}\supset\gamma_k\vee|\{\alpha_j\}|\;=\;1$
\end{enumerate}

\begin{proof} Consider first that
\begin{eqnarray}
P(\rho_{\{\alpha_j\}})=\sum_{\{\beta_m\}\in\{\alpha_j\}}P(\rho_{\{\beta_m\}})+P(\tr_{\{\beta_m\}\in\{\alpha_j\}}\rho)
\end{eqnarray}
and as well
\begin{eqnarray}
\sum_{m=2}^{j-1}\sum_{\{\epsilon_m\}\subset\{\alpha_j\}}\mathcal{E}(\rho_{\{\epsilon_m\}})\;\geq\;\sum_{\{\beta_m\}\in\{\alpha_j\}}P(\rho_{\{\beta_m\}})+P(\tr_{\{\beta_m\}\in\{\alpha_j\}}\rho)\\
\textrm{iff}\quad \exists\; \{\beta_m\}\in\{\alpha_j\}
\end{eqnarray}
such that the difference derives to
\begin{eqnarray}
P(\rho_{\{\alpha_j\}})-\sum_{m=2}^{j-1}\sum_{\{\epsilon_m\}\subset\{\alpha_j\}}\mathcal{E}_{\{\epsilon_m\}}(\rho)\;\leq\;0\\
\textrm{iff}\quad \exists \;\{\beta_m\}\in\{\alpha_j\}
\end{eqnarray}
from which consequently follows
\begin{eqnarray}
\mathcal{E}_{\{\alpha_j\}}(\rho)\;=\;0\qquad\forall\quad
\{\alpha_j\}\supset\gamma_k\quad\textrm{or}\quad|\{\alpha_j\}|\;=\;1
\end{eqnarray}
\end{proof}

\noindent\textbf{Property:}
\begin{enumerate}
    \item[P3:] $\mathcal{E}_{\{\alpha_j\}}(\rho^{\otimes n})\;=\;n\;\mathcal{E}_{\{\alpha_j\}}(\rho)$ (additivity on copies of the same state)
\end{enumerate}

\begin{proof}
Additivity follows directly if the conjecture is valid, i.e.
\begin{eqnarray}
\mathcal{E}_{\{\alpha_j\}}(\rho^{\otimes n})=\mathcal{E}_{\{\alpha_j\}}(\rho')
\end{eqnarray}
where the separability property of $\rho'$ is
\begin{eqnarray}
\gamma_k'=\{\{\gamma_k\}|\{\gamma_k\}|(\cdots)|\{\gamma_k\}\}
\end{eqnarray}
so the infimum is achieved in the appropriate decomposition such
that
\begin{eqnarray}
\mathcal{E}_{\{\alpha_j\}}(\rho^{\otimes
n})\;=\;n\;\mathcal{E}_{\{\alpha_j\}}(\rho)\;.
\end{eqnarray}
\end{proof}

\noindent\textbf{Property:}
\begin{enumerate}
    \item[P4:] $\mathcal{E}_{\{\alpha_j\}}(\lambda\; \rho_1 + (1-\lambda)\;\rho_2)\;\leq\;\lambda\; \mathcal{E}_{\{\alpha_j\}}(\rho_1)
+ (1-\lambda)\;\mathcal{E}_{\{\alpha_j\}}(\rho_2)\qquad$ (convexity)
\end{enumerate}

\begin{proof}
Invariance under local unitaries is easy to prove as the
constituting functions $P(\rho)$ are themselves invariant under
local unitary transformations
\begin{eqnarray}
P(\rho):=\inf_{p_i,\psi_i} \sum_i p_i (\sum_s S(\text{Tr}_{\neg
s}|\psi_i\rangle\langle\psi_i|))
\end{eqnarray}
and therefore
\begin{eqnarray}
\text{Tr}_{\neg s}(U_{\alpha_1} \otimes U_{\alpha_2} \otimes \cdots \otimes U_{\alpha_j}|\psi_i\rangle\langle\psi_i|(U_{\alpha_1} \otimes U_{\alpha_2}\otimes \cdots\otimes U_{\alpha_j})^\dagger)\\ =\text{Tr}_{\neg s}(\underbrace{(U_{\alpha_1} \otimes U_{\alpha_2} \otimes \cdots \otimes U_{\alpha_j})^\dagger U_{\alpha_1} \otimes U_{\alpha_2}\otimes \cdots\otimes U_{\alpha_j}}_{\id}|\psi_i\rangle\langle\psi_i|)
\end{eqnarray}
such that
\begin{eqnarray}
P(U_{\alpha_1} \otimes U_{\alpha_2} \otimes \cdots \otimes
U_{\alpha_j}\,\rho\, (U_{\alpha_1} \otimes U_{\alpha_2}\otimes
\cdots\otimes U_{\alpha_j})^\dagger)=P(\rho)\;.
\end{eqnarray}
\end{proof}

\noindent\textbf{Property:}
\begin{enumerate}
    \item[P5:] $\mathcal{E}_{\{\alpha_j\}}(\lambda\; \rho_1 + (1-\lambda)\;\rho_2)\;\leq\;\lambda\; \mathcal{E}_{\{\alpha_j\}}(\rho_1) + (1-\lambda)\;\mathcal{E}_{\{\alpha_j\}}(\rho_2)$ (Convexity)
\end{enumerate}

\begin{proof}
To prove that the total entanglement $E_{tot}(\rho)$ is convex one
needs to prove that  $P(\rho)$ is convex, which is trivial, as:
\begin{eqnarray}
&&P(\lambda\rho_1+(1-\lambda)\rho_2)\\&=&\inf_{p_i,r_j,\psi_i,\phi_j,\gamma_n,\tau_m}(\sum_i p_i \lambda \sum_s
S(\tr_{\neg_s}|\psi_i\rangle\langle\psi_i|)+\sum_j r_j (1-\lambda) \sum_s S(\tr_{\neg_s}|\phi_j\rangle\langle\phi_j|))\nonumber\\
&\leq&\underbrace{\inf_{p_i,\psi_i,\gamma_n}(\sum_i p_i \lambda
\sum_s S(\tr_{\neg_s}|\psi_i\rangle\langle\psi_i|))}_{\lambda
P(\rho_1)}+\underbrace{\inf_{r_j,\phi_j,\tau_m}(\sum_j r_j
(1-\lambda) \sum_s
S(\tr_{\neg_s}|\phi_j\rangle\langle\phi_j|))}_{(1-\lambda)P(\rho_2)}\nonumber
\end{eqnarray}
Note that the constituting elements $E_{\{\alpha_j\}}(\rho)$ are
only convex under local combinations of the form
\begin{eqnarray}
\rho\underbrace{+}_{local}\sigma:&=&\lambda\sum_i p_i(\rho^i_{\{\beta_1\}}\otimes\rho^i_{\{\beta_2\}}\otimes(\cdots)\rho^i_{\{\beta_k\}})+(1-\lambda)\sum_i p_i (\sigma^i_{\{\beta_1\}}\otimes\rho^i_{\{\beta_2\}}\otimes(\cdots)\rho^i_{\{\beta_k\}})\nonumber\\
&=&\sum_i p_i((\lambda\rho^i_{\{\beta_1\}}+(1-\lambda)\sigma^i_{\{\beta_1\}})\otimes\rho^i_{\{\beta_2\}}\otimes(\cdots)\rho^i_{\{\beta_k\}})
\end{eqnarray}
\end{proof}

\noindent\textbf{Property:}
\begin{enumerate}
   \item[P6:] $\sum_i \tr \left( V_i \rho V_i^\dag \right) \mathcal{E}_{tot} \left( \frac{V_i \rho V_i^\dag}{\tr \left( V_i \rho V_i^\dag \right)} \right) \leq \mathcal{E}_{tot}(\rho)$ (non-increasing on average under LOCC),\\
\end{enumerate}

\begin{proof}
This has already been proven, see e.g. Ref.~\cite{Plenio}.
\end{proof}

\bibliography{refsem}

\end{document}